\newcounter{myctr}

\documentclass{article}
\usepackage{url}
\usepackage{color}
\usepackage{enumerate}
\usepackage{amsmath}
\usepackage{amssymb}

\newcommand{\eqdef}{\stackrel{\mathrm{def}}{=}} % by definition
\def\be{\begin{equation}}
\def\ee{\end{equation}}

\def\bearn{\begin{eqnarray*}}
\def\eearn{\end{eqnarray*}}
\def\bear{\begin{eqnarray}}
\def\eear{\end{eqnarray}}
\def\barr{\begin{array}}
\def\earr{\end{array}}
\newcommand{\cro}[1]{\langle#1\rangle}   % bracket for integrals
\newcommand{\pdert}[1]{\frac{\partial#1}{\partial t}} % partial derivative wr t
\newcommand{\eref}[1]{Eq.(\ref{#1})}
\begin{document}

\makeatletter
\def\@biblabel#1{[#1]}
\makeatother

\markboth{J. G\'omez-Serrano and J.-Y. Le Boudec}{Comment on ``Mixing beliefs among interacting agents''}

%\catchline{}{}{}{}{}
%
%%%%%%%%%%%%%%%%%%%%%%%%%%%%%%%%%%%%%%%%%%%%%%%%%%%%%%%%%%%%%%%%%%%%

\title{Comment on ``Mixing beliefs among interacting agents''%COMMENT ON ``MIXING BELIEFS AMONG INTERACTING AGENTS''
}

\author{%JAVIER G\'OMEZ-SERRANO
Javier G\'omez-Serrano\footnote{J. G\'omez-Serrano's research was done while being an exchange student at EPFL.}, %JEAN-YVES LE BOUDEC
Jean-Yves Le Boudec}

%\address{
%Instituto de Ciencias Matem\'aticas \\
%Consejo Superior de Investigaciones Cient\'ificas (ICMAT CSIC-UAM-UCM-UC3M) \\
%C/Nicol\'as Cabrera, 13-15, 28049 Madrid \\
%javier.gomez@icmat.es \footnote{J. G\'omez-Serrano's research was done while being an exchange student at EPFL.}}

%\author{JEAN-YVES LE BOUDEC}
%\address{Laboratory for Computer Communications and Applications 2 \\
%\'Ecole Polytechnique F\'ed\'erale de Lausanne \\
%CH-1015 Lausanne, Switzerland\\
%jean-yves.leboudec@epfl.ch}

\maketitle

%\begin{history}
%\received{(received date)}
%\revised{(revised date)}
%\accepted{(Day Month Year)}
%\comby{(xxxxxxxxxx)}
%\end{history}

\begin{abstract}
We comment on the derivation of the main equation in the bounded
confidence model of opinion dynamics. In the original work, the equation is derived using an ad-hoc counting method. We point that the original derivation does contain some small mistake. The mistake does not
have a large qualitative impact, but it reveals the danger of the ad-hoc counting method.
We show how a more systematic approach, which we call micro to macro, can avoid such mistakes, without adding any significant complexity.
\end{abstract}


\section{Introduction}

One of the most famous mathematical models of opinion dynamics is presented in \cite{deffuant2000mixing}, in which the agents have continuous opinions and adjust them based on random encounters with other agents given that their opinions are below a fixed threshold $d$. Using a Taylor approximation and an ad-hoc counting method, the authors are able to derive an equation for the evolution of the distribution of opinions for small values of $d$.

We show that the original derivation does contain some small mistake. The mistake does not affect severely the model, as the qualitative behavior remains unchanged, but some values like the speed of convergence or the $n$-th order moments are impacted. This reveals the danger of the ad-hoc counting method.

We then show how the equation can be derived in a safer way, using a more systematic approach, which we call ``micro to macro". It consists in explicitly modelling the microscopic system before going to the fluid limit. The key technical tool is the drift equation. We illustrate that this is not more complex than the ad-hoc approach, and can be more easily validated.

\section{The Model}
\label{sec-model}
We consider a population of $N$ peers, each of them having an opinion $x_i \in [0,1]$ about some common subject. At every time step $k \in \mathbb{N}$, two peers are selected uniformly at random. If the distance between their opinions is sufficiently small, each of the opinions is modified, whereas if the opinions are far away they don't change. This is controlled with the uncertainty parameter $d$. Mathematically, we have that:

$\bullet$ If $|x_i - x_j| > d$ the opinions remain unchanged.

$\bullet$ If $|x_i - x_j| \leq d$ the opinions change the following way:
\vskip -0.cm
\begin{eqnarray*}
& x_i(k+1)-x_i(k) = \mu \cdot (x_j(k) - x_i(k)) & \\
& x_j(k+1) - x_j(k) = \mu \cdot (x_i(k) - x_j(k)) & \\
\end{eqnarray*}

where $\mu$ is a parameter (adaptation capacity) between 0 and $\frac{1}{2}$.

\section{The Ad-Hoc Counting Method in \cite{deffuant2000mixing,neau2000revisions}}
The final result is an equation for $\rho(t,x)$ defined as the PDF of opinions at time $t$:
 \be
\frac{\partial  \rho(x,t)}{\partial t}
= I_1+I_2
\label{eq-pdf0}
 \ee where $I_1$ and $I_2$ are described next. In \cite{deffuant2000mixing} only the final result is stated
without proof; the proof is available in
\cite{neau2000revisions}, but in French only. Therefore, in the
rest of this section we give a literal translation into English
of the relevant parts of \cite{neau2000revisions}.

\it
A simple enough calculation shows that, at the limit of the small values of $d$, the system has the tendency to amplify the irregularities of the population density according to their opinions, $\rho(x)$. During an elementary timestep corresponding to one interaction, the variation of $\rho(x)$ can be seen as the sum of two contributions:

A negative contribution corresponding to the probability for an agent of opinion $x$ to interact and modify his opinion:

\begin{equation}
 I_1 = -\rho(x) \int_{-d}^{d}\rho(x)\rho(x+y) dy \label{eq-1}
\end{equation}

A positive contribution $I_2$ corresponding to the probability that an agent of an initially different opinion has opinion $x$ after the interaction.

For this term, the obtaining of the normalizing constant is less trivial. The solution I have used is to consider a model in which the displacement of $x + y$ while interacting with $x + z$ is done around $\mu(y-z)$ with a gaussian probability. Letting the length of the gaussian go to zero, one gets our initial model conveniently normalized. The positive term is therefore calculated by the double integral:

\begin{eqnarray}
& \displaystyle I_2 = + \int dy \cdot \rho(x+y) \int dz \cdot \rho(x+z)\delta((x+y) + \mu((x+z)-(x+y)) = x) & \nonumber\\
 & \displaystyle = + \int dy \cdot \rho(x+y) \int dz \cdot \rho(x+z)\delta(y + \mu(z-y)) & \nonumber\\
 & \displaystyle = + \int dy \cdot \rho(x+y) \int dz \cdot \rho(x+z)\frac{1}{\mu}\delta(z + \frac{1-\mu}{\mu}y)&  \nonumber\\
 &\displaystyle = \frac{1}{\mu} \int_{-\mu d}^{\mu d} \rho(x+y) \cdot \rho (x + \frac{\mu-1}{\mu}y)dy &
\end{eqnarray}

For small $d$, one can replace in the integrals the terms in $\rho$ by their Taylor series around $x$ (limiting to second order in $y$):

\begin{eqnarray}
\label{taylor_integral}
& \displaystyle \delta \rho = -\int_{-d}^{d}\rho \cdot \left(\rho + y \rho' + \frac{y^2}{2}\rho''\right)dy & \nonumber \\
& \displaystyle + \frac{1}{\mu}\int_{-\mu d}^{\mu d}\left(\rho + y \rho' + \frac{y^2}{2}\rho''\right) \cdot \left(
\rho + \frac{\mu - 1}{\mu}y \rho' + \frac{1}{2}\left(\frac{\mu - 1}{\mu}y\right)^2\rho''\right)dy &
\end{eqnarray}
where all the terms in $\rho$ are evaluated at $x$ and $'$ represents the partial derivation with respect to $x$. Continuing the calculation, one obtains:

\begin{eqnarray}
\label{taylor_evaluated}
& \displaystyle \delta \rho = -2d\rho^2 - \frac{d^3}{3}\rho'' & \nonumber\\
& \displaystyle + \frac{1}{\mu}(2\mu d)\rho^2 + \frac{1}{\mu}\left[\frac{\mu - 1}{\mu}\rho'^2+\frac{1}{2}\left(\left(\frac{\mu - 1}{\mu}\right)^2+1\right)\rho \rho''\right]\cdot 2 \cdot \frac{(\mu d)^3}{3}&
\end{eqnarray}

After simplification, and supposing that the characteristic interaction time is $\tau$, one finally finds at the first non-zero order:

\begin{equation}
\label{final_result}
\frac{\partial \rho}{\partial t} = \frac{d^3}{2\tau}\cdot \mu \cdot (\mu - 1) \cdot \frac{\partial^2(\rho^2)}{\partial x^2}
\end{equation}
\normalfont
%\subsubsection{Corrections}
There are three errors that alter the final result published in \cite{deffuant2000mixing}.

\begin{enumerate}[(i)]
\item Equation (\ref{eq-1}) contains an extra term $\rho(x)$ and equation \eqref{taylor_evaluated} lacks a factor $\rho$ in the second term.

\item In both $I_1$ and $I_2$ there is a multiplying factor 2 missing. Correcting both mistakes $I_1$ should read:

\begin{equation}
 I_1 = -2\rho(x) \int_{-d}^{d}\rho(x+y) dy
\end{equation}

and $I_2$:

\begin{equation}
I_2 = \frac{2}{\mu} \int_{-\mu d}^{\mu d}\rho(x+y) \rho\left(x + \frac{\mu - 1}{\mu}y\right)dy
\end{equation}

%\item In the step between equations \eqref{taylor_evaluated} and \eqref{final_result}, when adding up the different terms, the final equation should read, after the correction of errors (i) and (ii):

\begin{equation}
\frac{\partial \rho}{\partial t} = \frac{2d^3}{3\tau}\cdot \mu \cdot (\mu - 1) \cdot \frac{\partial^2(\rho^2)}{\partial x^2}
\end{equation}

The error here is a division by 3/2 instead of 2. Note that
the error is propagated to \cite{weisbuch2003interacting} and, to the best of our knowledge, has not been corrected anywhere else.

\end{enumerate}

It is clear that the first correction is a typo, as it is corrected afterwards and the error is not propagated. Regarding correction (iii), it is a calculation mistake. Although the system's qualitative behavior isn't affected by this one, it might change the convergence speed of the system.
Correction (ii)
affects the time scale of the model by a factor of $2$%
\footnote{The system is described by \textit{``A chaque pas de temps, deux agents sont tir$\acute{e}$s au hasard dans l'ensemble de la population''} (At every timestep, two agents are randomly chosen from the population), which differs from what is computed in the sequel: the variation of $\rho(x)$ should be calculated as the positive and negative contribution (which are correctly written) for the case when the peer with opinion $x$ is the first element of the randomly chosen pair, and again, the same contribution (because of the symmetry of the problem) for the case when he is the second element of the randomly chosen pair.}. This is a common difficulty in the ad-hoc counting approach, which can be circumvented if, as we discuss next, we specify the microscopic model in detail.

%Correction (ii) is more serious because it affects the model. The system is described as \textit{``A chaque pas de temps, deux agents sont tir$\acute{e}$s au hasard dans l'ensemble de la population''} (At every timestep, two agents are randomly chosen from the population), which is opposed as what is afterwards calculated: the variation of $\rho(x)$ should be calculated as the positive and negative contribution (which are correctly written) for the case when the peer with opinion $x$ is the first element of the randomly chosen pair, and again, the same contribution (because of the symmetry of the problem) for the case when he is the second element of the randomly chosen pair. This is an informal justification about the incorrectness of the modeling.

The equation should read then,
following the notation from \cite{neau2000revisions}:

\begin{equation}
\frac{\partial  \rho(x,t)}{\partial t}
=
\frac{2}{\mu}\int_{x-\mu d}^{x+\mu d}\rho\biggl(\frac{x-(1-\mu)y}{\mu},t\biggr)\rho(y,t)\,dy
-2\rho(x,t)\int_{x-d}^{x+d}\rho(y,t)\,dy\,.
\label{eq-pdf}
\end{equation}

or, in a more compact (symmetric) way, as in a Boltzmann-like equation \cite{gressman2010global,villani2002review}, for $\mu \neq \frac{1}{2}$:

\begin{multline}
\label{eq-pdf-bis} \frac{\partial \rho(x,t)}{\partial t} =
\frac{2}{2\mu-1}\int_{x-d (2\mu-1)}^{x+d (2\mu-1)}
\rho\biggl(\frac{\mu x - (1-\mu)y}{2\mu-1},t\biggr)\rho\biggl(\frac{\mu y -
(1-\mu)x}{2\mu-1},t\biggr)\,dy
\\
-2\rho(x,t)\int_{x-d}^{x+d}\rho(y,t)\,dy
\end{multline}

%\subsection{Variance}
Note that this mistake has also propagated to \cite{ben2003bifurcations}, in which it affects the calculation of the evolution of the second order moments, as it should read $M_2(t) = M_2(0)e^{-M_0t}$ (there is a factor 2 in the exponent missing), and from there to several papers such as \cite{castellano2009statistical} or \cite{ben2003unity}. However, the derivation is also cited in \cite{lorenz2007continuous}, where it is partly corrected.
%\color{red} (there is a typo of a similar nature as (i) in equation (3) of the paper, where a factor $P(x,t)$ is missing).
%\color{black}

\section{An Alternative Method Based on Micro-to-Macro}
In this section we describe how the evolution equation can be obtained in a safer way, using a micro to macro approach. The idea \cite{benaim2008class} is to first write the ``drift equation" for the microscopic system, i.e. for the discrete time system defined in Section~\ref{sec-model} where the number of peers is finite. Then, in a second step, the evolution equation for densities as in Eq. (\ref{eq-pdf}) is obtained in the limit of infinitely many peers.

In the microscopic system, the density is not a proper one, but a mixture of Dirac masses. Instead of manipulating densities, it is more convenient (and faster) to manipulate the integral of arbitrary test functions against the density. Therefore, we proceed as follows.

%It is more convenient to handle the expectation of the density against drift equation isis is before writing the equation for
%In this section we will follow the same notation as in , \cite{1555363}. \color{blue} ADD MORE REFERENCES? \color{black} We begin taking as time unit $\frac{1}{N-1}$: this will ensure that the rate with which a peer makes a transition per time unit is constant for any value of $N$.

\paragraph{Step 1 (Micro)} Let $M^N(t)$ be the occupancy measure of the microscopic system with $N$ peers, defined as  $M^N(t) = \frac{1}{N}\sum_{n=1}^{N}\delta_{x_n^N(t)}$.
Let $h$ be any bounded, measurable function defined on the space of one peer, i.e. on $[0,1]$. The ``drift" of the microscopic system is the operator $\mathcal{G}^N$ defined by
$$ \mathcal{G}^N(h)(\nu) \eqdef \mathbb{E}\left(\left.h\left(M^N\left(t+1\right)\right)-h(M^N(t))\right|M^N(t) = \nu \right)$$
where the starting condition $\nu$ must be of the form $\nu = \frac{1}{N}\sum_{n = 1}^{N}\delta_{x_n}$. With some elementary manipulations we obtain:
\bear
 \lefteqn{\mathcal{G}^N(h)(\nu) =}
 \nonumber \\
 && \frac{2}{N(N-1)}\frac{1}{N}\sum_{m < n}(h(\mu x_m + (1-\mu)x_n) + h(\mu x_n + (1-\mu)x_m) - h(x_m) - h(x_n))1_{\{|x_n - x_m| \leq d\}}
  \nonumber\\
&=& \frac{1}{N^2(N-1)}\sum_{m < n}(h(\mu x_m + (1-\mu)x_n) + h(\mu x_n + (1-\mu)x_m) - h(x_m) - h(x_n))1_{\{|x_n - x_m| \leq d\}}
 \nonumber\\
&& + \frac{1}{N^2(N-1)} \sum_{n < m}(h(\mu x_m + (1-\mu)x_n) + h(\mu x_n + (1-\mu)x_m) - h(x_m) - h(x_n))1_{\{|x_n - x_m| \leq d\}}
  \nonumber\\
&=& \frac{1}{N^2(N-1)}\sum_{n,m}(h(\mu x_m + (1-\mu)x_n) + h(\mu x_n + (1-\mu)x_m) - h(x_m) - h(x_n))1_{\{|x_n - x_m| \leq d\}}
  \nonumber\\
&=& \frac{1}{N-1}\int_{[0,1]^2}(h(\mu y + (1-\mu)x) + h(\mu x + (1-\mu)y) - h(x) - h(y))1_{\{|x - y| \leq d\}}d\nu(x)d\nu(y)
 \nonumber
 \\
&=& \frac{2}{N-1}\int_{[0,1]^2}(h(\mu x+(1-\mu)y) - h(x))1_{\{|x - y| \leq d\}}d\nu(x)d\nu(y)\label{eq-calgl}
\eear

\paragraph{Step 2 (Micro to Macro)} Assume that when the number of peers $N$ grows towards $\infty$ the occupancy measure $M^N$ has a limit, say $\nu_t$. We must also re-scale time by a factor of $N$, as we see next, so that $M^N(k)\approx \nu_{k/N}$. Whether this holds, and why, is an entirely different problem, which is not addressed in \cite{deffuant2000mixing,neau2000revisions} and which do not address either here\footnote{If the state space is replaced by a discrete approximation, this is a classical result of \cite{ethier-kurtz-05}; for the model in Section~\ref{sec-model}, this is proved in \cite{gomez2010bounded}.}. Our purpose in this letter is solely to obtain a safe method for writing the equation. The method in e.g \cite{benaim2008class} says that, if we re-scale time by $1/N$, then the limit must satisfy
\bearn
 \pdert{\cro{h,\nu_t}}&= &\lim_{N\to\infty}N\mathcal{G}^N(h)(\nu_t)
 \eearn if the limit exists. Intuitively, this is because in the limit we have $\cro{h,\nu_{t+\frac{1}{N}}}-\cro{h,\nu_{t}}\approx
\mathcal{G}^N(h)(\nu_t)$.
By Eq.(\ref{eq-calgl}), we obtain:
  \bear
 \pdert{\cro{h,\nu_t}}= 2\int_{[0,1]^2}(h(\mu x+(1-\mu)y) - h(x))1_{\{|x - y| \leq d\}}d\nu(x)d\nu(y)\label{eq-calg2}
 \eear which is the main equation. To derive an equation for the density, we write $\nu_t(dx)=\rho(t,x) dx$, $\nu_t(dy)=\rho(t,y) dy$ and re-arrange \eref{eq-calg2} to obtain
 \bearn
 \int_{[0,1]}h(x) \pdert{\rho(t,x)}dx
 &=& 2\int_{[0,1]} h(x)\left[
\frac{1}{\mu}\int_{x-\mu d}^{x+\mu d}\rho\biggl(\frac{x-(1-\mu)y}{\mu},t\biggr)\rho(y,t)\,dy
 \right.
 \\
 &&\left.
 -\rho(x,t)\int_{x-d}^{x+d}\rho(y,t)\,dy\,
 \right] dx
 \eearn
\eref{eq-calgl} follows by identification of $\pdert{\rho(t,x)}$ to the term between square brackets.

\section{Conclusion}

We pointed out that there were some mistakes in the derivation of
Equation \eqref{final_result} in \cite{deffuant2000mixing}.
 The mistakes do not affect the
qualitative behavior of the system and were partly corrected in later works, however, their existence suggests that the ad-hoc counting method for deriving such equations is error prone. We described how a more systematic method can be used to avoid such pitfalls.

\appendix

\bibliographystyle{plain}
\bibliography{ws-acs}

\begin{tabular}{ll}
\textbf{Javier G\'omez-Serrano} & \textbf{Jean-Yves Le Boudec}\\
{\small Instituto de Ciencias Matem\'aticas} & {\small Laboratory for Computer Communications} \\
{\small Consejo Superior de Investigaciones Cient\'ificas} & {\small and Applications 2}\\
{\small C/ Nicol\'{a}s Cabrera, 13-15} & {\small \'Ecole Polytechnique F\'ed\'erale de Lausanne}\\
{\small Campus Cantoblanco UAM, 28049 Madrid} & {\small CH-1015 Lausanne, Switzerland}\\
{\small Email: javier.gomez@icmat.es} & {\small Email: jean-yves.leboudec@epfl.ch}\\
\end{tabular}

\end{document}